\documentclass[twocolumn,final]{elsarticle}

\usepackage{lineno,hyperref}
\usepackage{mhchem}
\modulolinenumbers[5]

\journal{Journal of Molecular Spectroscopy}

\newcommand*{\wn}{cm$^{-1}$}
\newcommand*{\AX}{$A^1\Pi-X^1\Sigma^+$}
\newcommand*{\dX}{$d^3\Delta-X^1\Sigma^+$}
\newcommand*{\eX}{$e^3\Sigma^--X^1\Sigma^+$}
\newcommand*{\aX}{$a'^3\Sigma^+-X^1\Sigma^+$}
\newcommand*{\DX}{$D^1\Delta-X^1\Sigma^+$}
\newcommand*{\IX}{$I^1\Sigma^--X^1\Sigma^+$}

\newcommand*{\Be}{$B^1\Sigma^+-e^3\Sigma^-$}
\newcommand*{\Bd}{$B^1\Sigma^+-d^3\Delta$}
\newcommand*{\BX}{$B^1\Sigma^+-X^1\Sigma^+$}

\newcommand*{\ds}{$d^3\Delta$}
\newcommand*{\es}{$e^3\Sigma^-$}

\newcommand*{\Ds}{$D^1\Delta$}

\newcommand*{\As}{$A^1\Pi$}

\newcommand*{\Xs}{$X^1\Sigma^+$}
\newcommand*{\0}{$v=0$}
\newcommand*{\1}{$v=1$}

\newcommand*{\4}{$v=4$}

\newcommand*{\CO}{$^{12}$C$^{16}$O}

\newcommand*{\change}[1]{#1}
\usepackage{verbatim}

\def\apj{Astroph.\  J.\ }
\def\apjs{Astroph.\  J.\ Suppl.\ Ser.\ }

\def\jcp{J. Chem.\ Phys.\ }
\def\cjp{Can.\ J.\ Phys.\ }

\def\jqsrt{J.\ Quant.\ Spectr.\ Rad.\ Transfer }
\def\pra{Phys.\ Rev.\ A }
\def\prl{Phys.\ Rev.\ Lett.\ }

\def\cp{Chem.\ Phys.\ }
\def\np{Nat.\ Phot.\ }

\def\jms{J. Mol.\ Spectrosc.\ }

\def\cp{Chem.\ Phys.\ }
\def\molp{Mol.\ Phys.\ }

\def\jpca{J. Phys.\ Chem.\ A}
\def\rsi{Rev. Scient.\ Instr. }

\def\aaa{Astron.\ Astrophys.\  }

\begin{document}

\begin{frontmatter}

\title{VUV-synchrotron absorption studies of N$_2$ and CO at 900 K}

\author[vua]{M. L. Niu}
\author[lei]{A. N. Heays}
\author[lei,not]{S. Jones}
\author[vua,usc]{E. J. Salumbides}
\author[lei]{E. F. van Dishoeck}
\author[sol]{\\N. De Oliveira}
\author[sol]{L. Nahon}
\author[vua]{W. Ubachs\corref{mycorrespondingauthor}}
\cortext[mycorrespondingauthor]{Corresponding author}
\ead{w.m.g.ubachs@vu.nl}

\address[vua]{Department of Physics and Astronomy, LaserLaB, VU University, De Boelelaan 1081, 1081 HV Amsterdam, The Netherlands}
\address[lei]{Leiden Observatory, Leiden University, PO Box 9513, 2300 RA Leiden, The Netherlands}
\address[not]{School of Physics and Astronomy, The University of Nottingham, University Park, Nottingham, NG7 2RD, United Kingdom}
\address[usc]{Department of Physics, University of San Carlos, Cebu City 6000, Philippines}
\address[sol]{Synchrotron Soleil, Orme des Merisiers, St. Aubin BP 48, 91192, Gif sur Yvette cedex, France}

\begin{abstract}
Photoabsorption spectra of \ce{N2} and CO were recorded at 900\,K,
using the vacuum-ultraviolet Fourier-transform spectrometer at the DESIRS beamline of synchrotron SOLEIL.
These high-temperature \change{and} high-resolution measurements allow for precise determination of line wavelengths, oscillator strengths, and predissociative line broadening of highly-excited rotational states with $J$ up to about 50, and also vibrational hot bands.
In CO, the perturbation of $A\,{}^1\Pi-X\,{}^1\Sigma^+$ vibrational bands $(0,0)$ and $(1,0)$ were studied, as well as the transitions to perturbing optically-forbidden states $e\,{}^3\Sigma^-$, $d\,{}^3\Delta$, $D\,{}^1\Delta$ and $a'\,{}^3\Sigma^+$.
In \ce{N2}, we observed line shifts and broadening in several $b\,{}^1\Pi_u-X\,{}^1\Sigma^+_g$ bands due to unobserved forbidden states of ${}^3\Pi_u$ symmetry.
The observed state interactions are deperturbed and, for \ce{N2}, used to validate a coupled-channels model of the interacting electronic states.
This data is appropriate for use in astrophysical or (exo-)planetary atmospheric applications where high temperatures are important and in future spectroscopic models of these molecules.
\end{abstract}


\end{frontmatter}


\section{Introduction}

\begin{figure}
\begin{center}
\includegraphics[width=\linewidth]{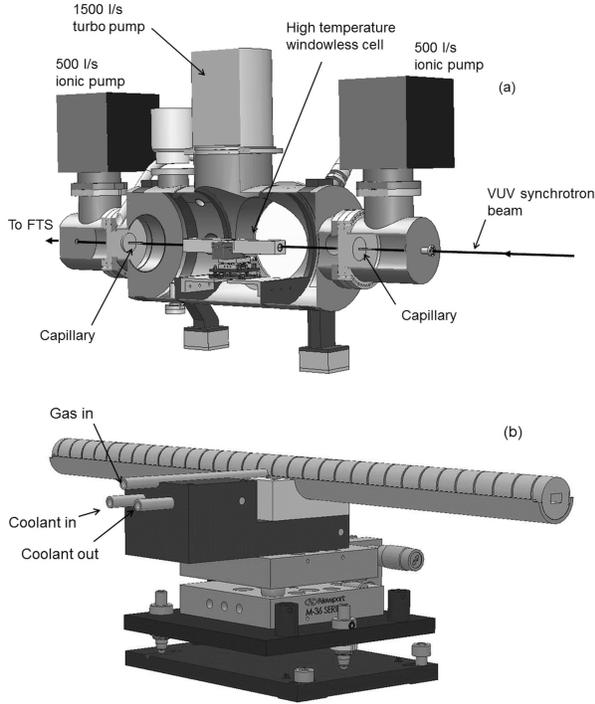}
\end{center}
\caption{(a) Cross-section drawing of the gas sample chamber mounted in the FTS-branch of the DESIRS beam line at SOLEIL. The high temperature windowless cell is located in the center of the chamber and is separated from the ultra-high vacuum of the beamline by two stages of differential pumping. (b) The inset shows details of the cell and the shielding where half of the cylindrical shell has been removed for clarity. The heating element is wrapped all over the cylindrical cell inside a groove in order to increase surface contact with the cell. The gas is flowing through a $7.5 \times 4.5$ mm  tube into the heated cell. The copper base can be cooled down with a thermalized water circulation system.}
\label{fig:hotcell}
\end{figure}

The technique of Fourier-transform spectroscopy is typically applied to the infrared and optical wavelength domains. The interferometric principle requires a beam-splitter for which no materials exist in the far vacuum ultraviolet (VUV) part of the electromagnetic spectrum. At the DESIRS beamline of the SOLEIL synchrotron~\cite{nahon2012} this problem was solved by developing a VUV Fourier-transform spectrometer (FTS) based on beam-splitting by wave-front division, thus enabling high-resolution spectroscopy at wavelengths in the range $40-200$\,nm~\cite{oliveira2009,oliveira2011}. In recent years, this unique instrument has been used to perform high resolution spectroscopic studies on a number of small molecules in the gas phase that exhibit strongly-structured multi-line spectra, such as H$_2$~\cite{dickenson2011}, HD~\cite{ivanov2010}, N$_2$~\cite{heays2011b}, and CO~\cite{niu2013}; as well as for molecules with more continuum-like spectra, such as CO$_2$~\cite{archer2013}. These studies amply demonstrate the multiplex advantage of the Fourier-transform technique by revealing many hundreds of absorption lines in a single-scan window of some 5\,nm, determined by the bandwidth of the beam line undulator source. Alternatively, the setup was used to determine photo-absorption cross sections~\cite{stark2014} and predissociation linewidths (and hence rates) of excited states of small molecules~\cite{heays2011b,eidelsberg2012}.

The FTS-VUV setup has been used for gas-phase absorption spectroscopy under varied measurement conditions. Most studies have been performed in a quasi-static gas environment where the gas sample  effusively flows through a narrow capillary-shaped absorption cell, with the absorption path aligned with the VUV beam emanating from the undulator. This cell was not equipped with windows, to permit passage of the VUV beam through the sample gas into the FTS-VUV instrument for spectroscopic analysis. For this geometry, differential pumping maintains an ultrahigh vacuum in the FTS and DESIRS beam line. The column density of absorbing gas is limited by the pumping conditions and vacuum requirements of the beam line. In any case, there is a pressure gradient over the cell length falling off toward both ends, complicating any absolute column density calibration.
In further studies dedicated to cross-section measurements, a movable gas cell was used of $\sim 19$\,mm length and sealed by either MgF$_2$ or LiF wedged windows. This allowed for somewhat-higher pressures and controlled gas column densities~\cite{gavilan2013}. Studies using this cell are limited in wavelength range to $\lambda > 105$ nm~\cite{federmann2001} by the short-wavelength opacity of the windows.
In some experiments requiring the simplification of congested spectra, a molecular jet expansion was employed as well as cooling of the quasi-static gas cell with liquid-nitrogen or liquid-helium. A comparison between these techniques was performed in a study of the D$_2$ spectrum~\cite{delange2012} which also demonstrated improved spectral resolution and accuracy through reduction of the Doppler width under these conditions.

For the present study a heated cell is implemented, allowing for the recording of spectra at temperatures of $\sim 1000$\,K. The high-resolution FTS allows for the measurement and analysis of severely congested spectra at these elevated temperatures. Such spectra bear significance for the modeling of astrophysical shock-wave regions~\cite{heays2014}, or other high-temperature astrophysical regions where the spectroscopy of small molecules is key to understanding the phenomena, such as e.g., in the photospheres of white dwarfs~\cite{bagdonaite2014}. Another goal of performing spectroscopy of hot samples is to follow rotational progressions to high $J$-quantum numbers, where perturbations due to non-Born-Oppenheimer effects are abundantly present. Some pertinent perturbation features specifically occurring at high rotational quantum numbers will be shown here in VUV-absorption spectra of CO and N$_2$ recorded at 900\,K.

\section{Experimental}
\label{sec:experimental}

The VUV Fourier-Transform spectrometer at the DESIRS beamline is a permanent end station dedicated to high-resolution photoabsorption studies in the range $4-30$ eV~\cite{nahon2012}. The instrument has been described in detail previously~\cite{oliveira2009,oliveira2011}. In short, the spectrometer is based on wave-front division interferometry using reflective surfaces, thus allowing the extension of the FTS technique into the far VUV spectral range. The undulator white beam is used as the background, feeding the FTS branch and permits the recording of a spectral bandwidth $\Delta E/E =7$\%, corresponding to 5 nm, on each scan.  The typical integration time for a single scan is less than 30 minutes to obtain a signal-to-noise ratio for the background continuum level of $\sim 400$.

The windowless absorption cell is a 40 cm long T-shaped tube with a rectangular cross section, installed under vacuum inside the multi-purpose gas sample chamber of the FTS branch (Fig.~\ref{fig:hotcell}). The cross section of the tube (7.5  $\times$ 4.5 mm) is adapted to the astigmatic shape and dimensions of the undulator source in this section of the beam line. An Inconel heating element (thermocoax) is wrapped around the tube sitting in a groove designed to maximize the contact surface between the heating wire and the cell, and ensuring the gas flowing in the tube is uniformly heated. Two semi-cylindrical shells are pressed around the cell in order to improve the thermal contact during the heating operation. An extra stainless steel box is also installed to shield radiation originating from the cell. Inconel allows for heating the cell up to $1000$ K, although, the present measurements were done at a maximum temperature of $900$ K. The cell is mounted on a copper base plate that can be cooled by water circulation, although during the experiments the setup was operated without the cooling system. The temperature of the base-mount was carefully monitored within the covered range of temperature (for the cell) and never went beyond a maximum of $200^o$C with no visible consequence or damage. A thermocouple is connected at one end of the cell to obtain an indication of the gas temperature.  The quasi-static gas density inside the heated cell was monitored from outside the vacuum by a 1 mbar range capacitive gauge. The gas column density was adjusted by a needle valve in order to have a constant continuous flow through the cell during the photoabsorption measurements. The effective column density along the absorption path inside the cell varied from $4\times10^{14}$ to $1.2\times10^{17}$\,cm$^{-2}$ and was adjusted according to the cross sections of the recorded bands.

In the present study, the FTS-VUV was set to provide an instrumental linewidth of 0.27\,\wn. The Doppler broadening corresponds to 0.28 \wn\ at a frequency of 65\,000\,\wn, temperature of 900\,K, and molecular mass of 28\,amu. After convolution with the instrument width, a spectral linewidth of 0.39 \wn\ is anticipated for unsaturated and \change{non-predissociation-broadened} N$_2$ and CO lines.
The FTS spectra are intrinsically wavelength calibrated by monitoring the movement of the travel arm in the interferometer which is controlled by a HeNe-laser~\cite{oliveira2009,oliveira2011}. Additional and improved calibration can be derived from co-recording special calibration lines, e.g. resonance lines of noble gases~\cite{brandi2001,saloman2004}. In the present case of the CO spectra the very accurate laser-calibration data of the low-$J$ rotational lines in the $A-X$ bands, accurate to $\Delta\lambda/\lambda = 3 \times 10^{-8}$, were used~\cite{salumbides2012}. For the present high-temperature measurements with larger Doppler-broadening the FTS was not used in its very highest resolution mode and the spectral accuracy typically reached is estimated at 0.02 \wn. The accuracy is somewhat lower for weaker and blended lines.

\section{Absorption spectra of N$_2$}
\label{sec:N2absorption}

\begin{figure*}
  \centering
  \includegraphics{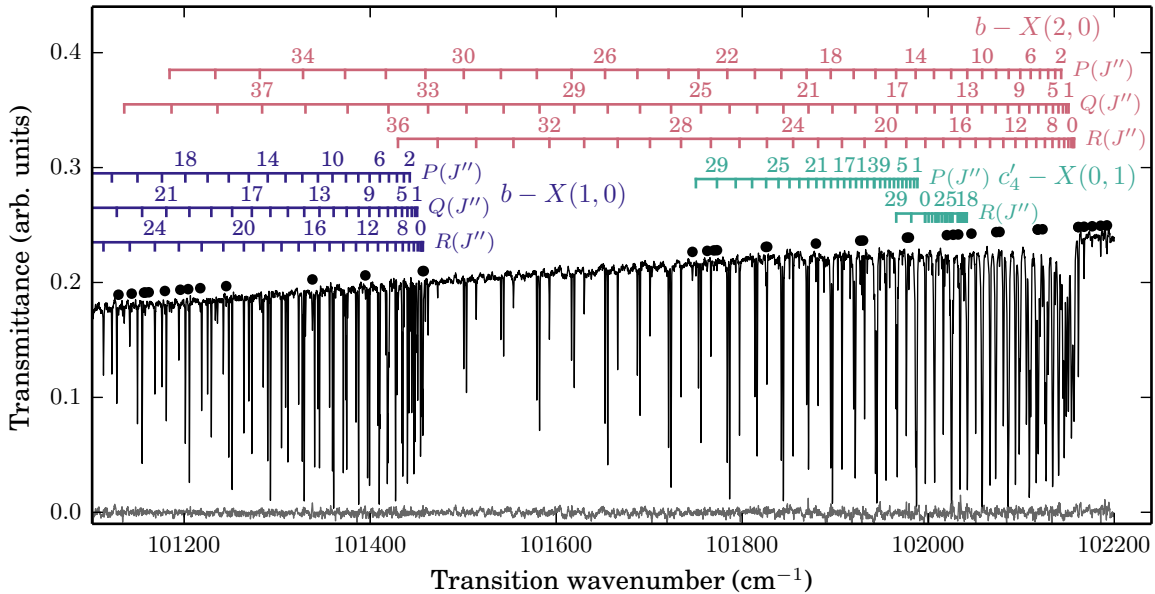}
  \caption{Photoabsorption spectrum showing the bands $b-X(2,0)$, $c'_4-X(0,1)$, and part of $b-X(1,0)$; and further absorption lines arising from H$_2$ contamination, high-$J'$ lines of $b-X(3,0)$, and of unknown origin (\emph{circles}). The lower trace indicates the residual error after subtracting a model spectrum.}
  \label{fig:scan N2 b02-X00}
\end{figure*}
Five N$_2$ vibrational bands were analysed appearing in our spectrum between 100\,400 and 108\,500\,cm$^{-1}$ (99.6 and 92.2\,nm).
These bands are spectroscopically denoted $b\,{}^1\Pi_u-X\,{}^1\Sigma_g^+(v',v''=0)$ for $v'=$ 0, 1, 2, and 10, and $c'_4{}^1\Sigma_u^+-X{}^1\Sigma_g^+(v'=0,v''=1)$ (where $v'$ and $v''$ are upper- and lower-state vibrational quantum numbers, respectively; and hereafter we neglect electronic-state term symbols); and have been previously observed in room-temperature or expansion-cooled synchrotron- or laser-based experiments~\cite{ubachs_etal1989,levelt1992,sprengers_etal2004b,sprengers_etal2005,stark_etal2005,stark_etal2008}.
Part of our photoabsorption spectrum showing three of these bands is plotted in Fig.~\ref{fig:scan N2 b02-X00}.
A listing of the deduced term values for the $e$ and $f$ components of
the observed $b(v')$ levels is given in Table~\ref{tab:N2 term values}.

\begin{table*}
  \begin{minipage}{1.0\linewidth}
    \caption{Experimental upper term values\protect\footnote{With units of cm$^{-1}$ and parenthetical $1\sigma$ fitting uncertainties in terms of the least-significant digit. The estimated absolute calibration uncertainty is 0.04\,cm$^{-1}$.} for observed lines in \ce{N2} indexed by excited state angular-momentum, $J'$.}
    \centering
    \scriptsize
    \label{tab:N2 term values}
    \begin{tabular}{clllllll}
      \hline\hline
      \\[-1ex]
           &\multicolumn{2}{c}{\large $b(0)$}   &\multicolumn{2}{c}{\large $b(1)$}   &\multicolumn{2}{c}{\large $b(2)$}   &\multicolumn{1}{c}{\large $b(10)$} \\
      $J'$ & \multicolumn{1}{c}{$e$-parity} & \multicolumn{1}{c}{$f$-parity} & \multicolumn{1}{c}{$e$-parity} & \multicolumn{1}{c}{$f$-parity} & \multicolumn{1}{c}{$e$-parity} & \multicolumn{1}{c}{$f$-parity} & \multicolumn{1}{c}{$e/f$-parity\footnote{No splitting of $e$- and $f$- parity levels was observed (apart from for $J'=18$) and these were assumed identical.}} \\
      \hline
      \\[-1ex]
      1  & 100\,819.84(2)         & 100\,819.91(4)         & 101\,454.460(5)        & 101\,454.455(8)        & 102\,154.82(1)         & 102\,154.79(3)      & 108\,374.115(7)                   \\
      2  & 100\,825.54(3)         & 100\,825.61(1)         & 101\,460.090(7)        & 101\,460.097(3)        & 102\,160.33(1)         & 102\,160.359(7)     & 108\,378.968(4)                   \\
      3  & 100\,834.25(1)         & 100\,834.27(2)         & 101\,468.560(2)        & 101\,468.541(4)        & 102\,168.691(5)        & 102\,168.658(9)     & 108\,386.239(4)                   \\
      4  & 100\,845.85(2)         & 100\,845.821(9)        & 101\,479.798(4)        & 101\,479.803(3)        & 102\,179.754(8)        & 102\,179.731(8)     & 108\,395.955(4)                   \\
      5  & 100\,860.303(7)        & 100\,860.32(1)         & 101\,493.874(2)        & 101\,493.877(3)        & 102\,193.637(4)        & 102\,193.70(1)      & 108\,408.050(2)                   \\
      6  & 100\,877.67(1)         & 100\,877.660(8)        & 101\,510.764(3)        & 101\,510.75(1)         & 102\,210.295(7)        & 102\,210.267(6)     & 108\,422.591(3)                   \\
      7  & 100\,897.887(6)        & 100\,897.895(9)        & 101\,530.443(2)        & 101\,530.456(4)        & 102\,229.655(4)        & 102\,229.675(8)     & 108\,439.530(2)                   \\
      8  & 100\,921.01(1)         & 100\,920.987(7)        & 101\,552.946(2)        & 101\,552.940(3)        & 102\,251.874(9)        & 102\,251.846(8)     & 108\,458.875(2)                   \\
      9  & 100\,946.968(6)        & 100\,946.96(1)         & 101\,578.238(2)        & 101\,578.24(1)         & 102\,276.746(6)        & 102\,276.770(9)     & 108\,480.618(2)                   \\
      10 & 100\,975.83(1)         & 100\,975.815(7)        & 101\,606.338(3)        & 101\,606.348(6)        & 102\,304.416(6)        & 102\,304.426(7)     & 108\,504.758(2)                   \\
      11 & 101\,007.503(6)        & 101\,007.52(3)         & 101\,637.230(2)        & 101\,637.230(3)        & 102\,334.845(3)        & 102\,334.856(7)     & 108\,531.283(2)                   \\
      12 & 101\,042.036(9)        & 101\,042.04(1)         & 101\,670.911(2)        & 101\,670.912(3)        & 102\,368.018(4)        & 102\,368.018(7)     & 108\,560.191(2)                   \\
      13 & 101\,079.458(7)        & 101\,079.44(1)         & 101\,707.379(2)        & 101\,707.378(3)        & 102\,403.933(3)        & 102\,403.926(8)     & 108\,591.479(2)                   \\
      14 & 101\,119.70(1)         & 101\,119.690(8)        & 101\,746.629(2)        & 101\,746.626(3)        & 102\,442.572(3)        & 102\,442.635(8)     & 108\,625.157(2)                   \\
      15 & 101\,162.74(1)         & 101\,162.74(1)         & 101\,788.652(2)        & 101\,788.654(3)        & 102\,483.948(2)        & 102\,483.965(6)     & 108\,661.209(2)                   \\
      16 & 101\,208.58(2)         & 101\,208.600(9)        & 101\,833.455(2)        & 101\,833.456(3)        & 102\,528.044(3)        & 102\,528.086(5)     & 108\,699.644(2)                   \\
      17 & 101\,257.250(9)        & 101\,257.29(2)         & 101\,881.006(2)        & 101\,881.013(3)        & 102\,574.875(2)        & 102\,574.887(4)     & 108\,740.459(3)                   \\
      18 & 101\,308.77(2)         & 101\,308.71(1)         & 101\,931.329(3)        & 101\,931.332(3)        & 102\,624.387(3)        & 102\,624.415(4)     & 108\,783.684(5)                   \\
      19 & 101\,362.95(1)         & 101\,362.91(3)         & 101\,984.400(2)        & 101\,984.400(4)        & 102\,676.610(2)        & 102\,676.647(4)     & 108\,829.02(1)\footnote{A splitting of $e$- and $f$- parity levels was observed, with term values  108\,829.02(1) and 108\,828.82(3)\,cm$^{-1}$, respectively.}  \\
      20 & 101\,419.98(3)         & 101\,419.92(1)         & 102\,040.210(3)        & 102\,040.215(3)        & 102\,731.545(3)        & 102\,731.574(3)     & 108\,876.398(7)                   \\

      21 & 101\,479.69(2)         & 101\,479.66(4)         & 102\,098.767(2)        & 102\,098.764(5)        & 102\,789.153(2)        & 102\,789.168(4)     & 108\,926.565(6)                   \\
      22 & 101\,542.27(6)         & 101\,542.13(3)         & 102\,160.044(4)        & 102\,160.042(3)        & 102\,849.447(2)        & 102\,849.489(3)     & 108\,979.148(5)                   \\
      23 & 101\,607.38(2)         & 101\,607.27(6)         & 102\,224.043(3)        & 102\,224.045(6)        & 102\,912.413(2)        & 102\,912.457(4)     & 109\,034.005(5)                   \\
      24 & \multicolumn{1}{c}{--} & 101\,675.24(3)         & 102\,290.766(6)        & 102\,290.761(4)        & 102\,978.050(3)        & 102\,978.097(3)     & 109\,091.321(4)                   \\
      25 & 101\,745.90(5)         & 101\,745.66(7)         & 102\,360.177(4)        & 102\,360.171(8)        & 103\,046.349(2)        & 103\,046.384(4)     & 109\,150.906(5)                   \\
      26 & \multicolumn{1}{c}{--} & 101\,819.01(6)         & 102\,432.300(8)        & 102\,432.275(6)        & 103\,117.309(4)        & 103\,117.327(3)     & 109\,212.899(4)                   \\
      27 & 101\,895.03(5)         & \multicolumn{1}{c}{--} & 102\,507.081(5)        & 102\,507.08(1)         & 103\,190.873(2)        & 103\,190.911(5)     & 109\,277.088(7)                   \\
      28 & \multicolumn{1}{c}{--} & 101\,973.59(9)         & 102\,584.52(1)         & 102\,584.532(8)        & 103\,267.083(4)        & 103\,267.124(4)     & 109\,343.734(9)                   \\
      29 & \multicolumn{1}{c}{--} &\multicolumn{1}{c}{--}  & 102\,664.686(8)        & 102\,664.67(2)         & 103\,345.914(3)        & 103\,345.956(6)     & 109\,412.79(1)                    \\
      30 & \multicolumn{1}{c}{--} & 102\,138.44(6)         & 102\,747.45(4)         & 102\,747.45(1)         & 103\,427.360(6)        & 103\,427.403(5)     & 109\,484.23(4)                    \\
      31 & \multicolumn{1}{c}{--} &\multicolumn{1}{c}{--}  & 102\,832.91(1)         & 102\,832.83(3)         & 103\,511.408(4)        & 103\,511.446(8)   & \multicolumn{1}{c}{--}              \\
      32 & \multicolumn{1}{c}{--} &\multicolumn{1}{c}{--}  & \multicolumn{1}{c}{--} & 102\,920.94(2)         & 103\,598.05(1)         & 103\,598.083(6)   & \multicolumn{1}{c}{--}              \\
      33 & \multicolumn{1}{c}{--} &\multicolumn{1}{c}{--}  & 103\,011.62(3)         & \multicolumn{1}{c}{--} & 103\,687.265(6)        & 103\,687.32(1)    & \multicolumn{1}{c}{--}              \\
      34 & \multicolumn{1}{c}{--} &\multicolumn{1}{c}{--}  & \multicolumn{1}{c}{--} &\multicolumn{1}{c}{--}  & 103\,779.06(1)         & 103\,779.09(2)    & \multicolumn{1}{c}{--}              \\
      35 & \multicolumn{1}{c}{--} &\multicolumn{1}{c}{--}  & \multicolumn{1}{c}{--} &\multicolumn{1}{c}{--}  & 103\,873.41(1)         & 103\,873.46(2)    & \multicolumn{1}{c}{--}              \\
      36 & \multicolumn{1}{c}{--} &\multicolumn{1}{c}{--}  & \multicolumn{1}{c}{--} & 103\,299.14(5)         & 103\,970.27(3)         & 103\,970.33(2)    & \multicolumn{1}{c}{--}              \\
      37 & \multicolumn{1}{c}{--} &\multicolumn{1}{c}{--}  & \multicolumn{1}{c}{--} &\multicolumn{1}{c}{--}  & 104\,069.79(4)         & 104\,069.80(5)    & \multicolumn{1}{c}{--}              \\
      38 & \multicolumn{1}{c}{--} &\multicolumn{1}{c}{--}  & \multicolumn{1}{c}{--} &\multicolumn{1}{c}{--}  & \multicolumn{1}{c}{--} & 104\,171.73(4)    & \multicolumn{1}{c}{--}              \\
      39 & \multicolumn{1}{c}{--} &\multicolumn{1}{c}{--}  & \multicolumn{1}{c}{--} &\multicolumn{1}{c}{--}  & \multicolumn{1}{c}{--} & 104\,276.17(9)    & \multicolumn{1}{c}{--}              \\
      40 & \multicolumn{1}{c}{--} &\multicolumn{1}{c}{--}  & \multicolumn{1}{c}{--} &\multicolumn{1}{c}{--}  & \multicolumn{1}{c}{--} & 104\,383.05(5)    & \multicolumn{1}{c}{--}              \\
      \hline\hline
    \end{tabular}
  \end{minipage}
\end{table*}

The analysis of DESIRS FTS spectra of molecular nitrogen has been discussed previously~\cite{heays2011b}.
This involves simulating each observed absorption line with a Voigt profile defined by a Gaussian Doppler width, Lorentzian natural line width, transition wavenumber, and integrated cross section.
A summed cross section is then transformed into an absorption spectrum by the Beer-Lambert law and convolved with a sinc function simulating the instrumental resolution of the FTS.
All parameters defining the model absorption spectrum are then automatically optimised to best agree with the experimental scan.

In many cases a more useful measurement of the strength of a line than the integrated cross section is a derived band $f$-value, calculated by factoring the ground-state rotational thermal population as well as rotation-dependent H\"onl-London linestrength factors.
Band $f$-values are only weakly dependent on upper-state $J'$ for unperturbed bands.

The main difficulties encountered while analysing the hot-cell N$_2$ spectrum were the significant contamination from highly-excited rotational structure of nearby bands and obtaining a correct calibration of the temperature in the cell.
Groups of lines from the same vibrational band were sometimes analysed while assuming correlated wavenumbers, widths and strengths to facilitate the analysis of blended spectral regions.
That is, $P(J''-1)$ and $R(J''+1)$ transitions are connected to a common excited state so the difference in their transition wavenumbers was fixed to known ground-state energy levels~\citep{edwards_etal1993} and a common linewidth assumed.
A weak $J'$-dependence (or $J'$-independence) was also assumed for some linewidths or $f$-values.

Lines with natural widths below about 0.05\,cm$^{-1}$ full-width half-maximum (FWHM) are not reliably measured in our experiment due to concurrent instrument and Doppler broadening by 0.27 and about 0.4\,cm$^{-1}$\,FWHM, respectively.
No linewidths are then measurable from our spectrum for transitions to the weakly-predissociated $c'_4(0)$ level~\cite{ubachs_etal2001}.

We compare our measured $f$-values and linewidths with those calculated from an existing model of N$_2$ photoabsorption and dissociation, including photoabsorbing ${}^1\Pi_u$ and ${}^1\Sigma^+_u$ excited states and spin-forbidden but dissociative ${}^3\Pi_u$ states~\cite{heays2011b,lewis_etal2005a,haverd2005,lewis_etal2008b,heays2014b}.
This model solves a coupled-Schr\"odinger equation (CSE) for the nuclear motion of the excited molecule, where the necessary potential-energy curves and state interactions have been optimised with respect to a large body of room-temperature experimental data.
This model has been successfully employed previously in applications of atmospheric~\cite{lavvas_etal2011} and astronomical photochemistry~\cite{li2013,heays2014a}, including temperatures as high as 1000\,K.
Here, we seek to validate the extrapolation of the CSE model to high temperature by comparison to our new measurements.

\subsection{Temperature calibration}

\begin{figure}
  \centering
  \includegraphics{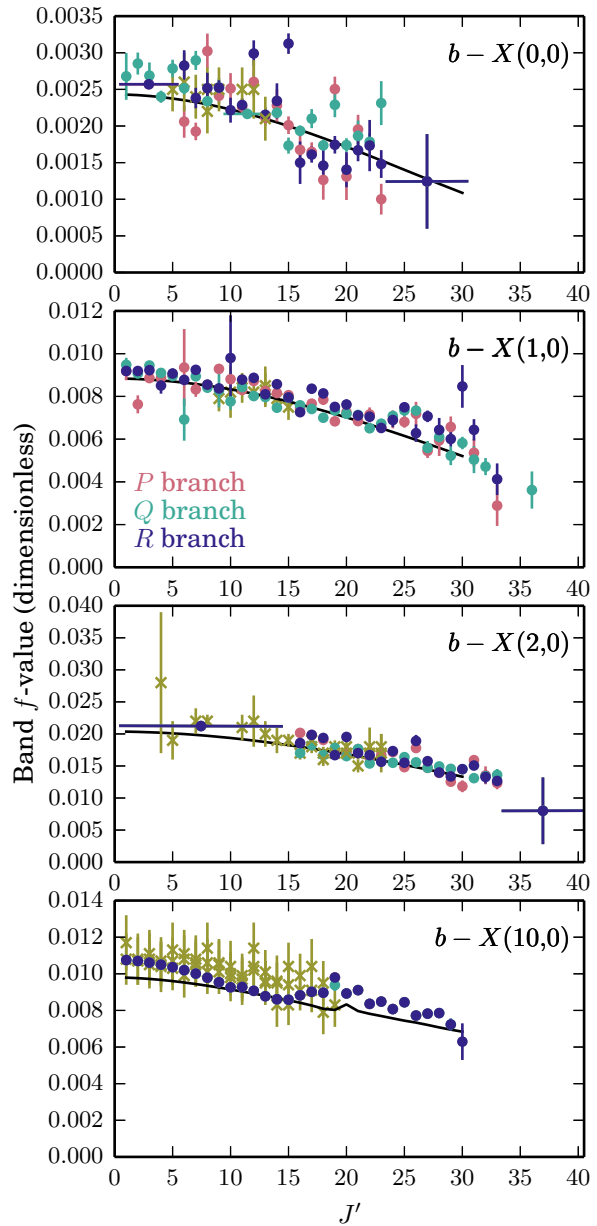}
  \caption{Band $f$-values of all transitions observed in our experiment as a function of excited-state angular-momentum quantum number, $J'$, and with $1\sigma$ random fitting uncertainties (\emph{circles with error bars}). A 10\% systematic error also applies and some $f$-values were analysed assuming $J'$-independent ranges (\emph{horizontal error bars}).
    Also shown are previously-measured $f$-values~\cite{stark_etal2005,stark_etal2008} (\emph{crosses})\change{, and calculated by the CSE model (\emph{solid black curves})}.
    }
  \label{fig:compare strengths}
\end{figure}
The $f$-values of $b-X(v',0)$ transitions were used to calibrate the ground-state rotational temperature and N$_2$ column density in the hot cell by comparison with previously-measured absolutely-calibrated $f$-values for $v'=0,1,2$, and 10~\cite{stark_etal2005,stark_etal2008}.
The resultant values are $(6.35 \pm 0.64)\times 10^{15}$\,cm$^{-2}$ and $901 \pm 26$\,K, respectively.
The reference data was recorded at room temperature, included rotational levels as high as $J=23$, and themselves have an absolute column density uncertainty of 10\% which is also the dominant systematic uncertainty of our $f$-values.
The final agreement between the present measurements and the reference data, shown in Fig.~\ref{fig:compare strengths}, is very good despite the factor-of-5 difference in ground state populations, for example, at $J'=20$.

\begin{figure}
  \centering
  \includegraphics{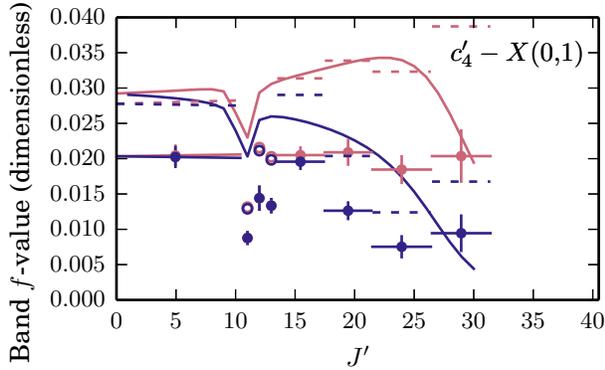}
  \caption{Band $f$-values of all $c'_4-X(0,1)$ transitions observed in our experiment as a function of excited-state angular-momentum quantum number, $J'$, and with $1\sigma$ random fitting uncertainties (\emph{circles with error bars}). A 10\% systematic error also applies and some $f$-values were analysed assuming $J'$-independent ranges (\emph{horizontal error bars}).
    Also shown are alternative experimental $f$-values assuming an 800\,K ground state excitation (\emph{dashed lines, open circles)} and reference values calculated from a combination of CSE and experimental data (\emph{solid curve}).
    }
  \label{fig:compare cp00-X01 strengths}
\end{figure}
The $c'_4-X(0,1)$ band appears quite weakly in our spectrum and was analysed in order to estimate the vibrational temperature in the hot cell.
For this, constant band $f$-values were assumed over small ranges of most $P(J'')$ and $R(J'')$ lines as indicated piecewise in Fig.~\ref{fig:compare cp00-X01 strengths}.
Simulated $c'_4-X(0,1)$ $f$-values are also shown, with magnitude calculated from the ratio of $c'_4-X(0,0)$ and $c'_4-X(0,1)$ $f$-values deduced by electron-excited fluorescence~\cite{liu_etal2008}, $f_{(0,0)}/f_{(0,1)}=6.3\pm 0.4$, and an absolute $c'_4-X(0,0)$ absorption $f$-value measurement~\cite{stark_etal2005}.
The stated uncertainties of the two experimental values used in this comparison are 6\%~\cite{liu_etal2008} and 10\%~\cite{stark_etal2005}, respectively, although the latter should be neglected because our experimental column-density is calibrated to the same reference.
We used the CSE model to simulate the significant rotational dependence of $c'_4- X(0,1)$ $f$-values and assumed a 900\,K distribution of ground-state rovibrational levels.
This simulation then correctly reproduced the observed splitting of $P$- and $R$-branch $f$-values for $c'_4- X(0,1)$ transitions with increasing $J'$.
This splitting is also known to occur for the $c'_4- X(0,0)$ fundamental band~\cite{stark_etal2005} and is the result of a rotational-perturbation of $c'_4(v'=0)$ by nearby ${}^1\Pi_u$ levels~\cite{heays2011b}.
Additionally, $c'_4- X(0,1)$ transitions to $J'=11$, 12, and 13 levels are significantly weakened relative to their neighbours due to a well-known localised perturbation by the crossing rotational term series of $b'\,{}^1\Sigma^+_u(v'=1)$~\cite{levelt1992}.

The newly-measured $c'_4-X(0,1)$ $f$-values are somewhat smaller than the simulated values
and an alternative model adopting an 800\,K distribution of ground-state levels leads to the better agreement indicated in Fig.~\ref{fig:compare cp00-X01 strengths}.
This may indicate incomplete thermalisation of the N$_2$ in our experiment leading to a lesser degree of vibrational excitation than rotational.
A similar result is found in Sec.~\ref{sec:COabsorption} for the CO rotational and vibrational temperatures.

As a final check on the temperature of our sample of N$_2$, the Doppler broadening in our experiment was measured by reference to lower-$J'$ levels of the $b-X(1,0)$ absorption band, whose predissociation broadening is known to be below our resolution limit~\cite{sprengers_etal2004b,lewis_etal2005a}.
We find a kinetic temperature from this of about 930\,K, with an uncertainty estimated to be significantly greater than for our deduced rotational temperature.

\subsection{Results}

Transition wavenumbers for all observed $b-X(v',0)$ bands were reduced to term values using accurate N$_2$ ground-state molecular constants~\cite{edwards_etal1993}.
Term values for these bands have been deduced previously for rotational levels with $J'$ as high as 36 and with about $0.1$\,cm$^{-1}$ uncertainty.
Our term values are listed in Table \ref{tab:N2 term values} and have statistical uncertainties of around 0.01\,cm$^{-1}$.
The absolute calibration of our experiment was made by comparison of argon resonance lines appearing in our spectra with the NIST database and has an estimated uncertainty of 0.04\,cm$^{-1}$.

Our deduced band $f$-values are plotted in Fig.~\ref{fig:compare strengths}.
The decrease of $b-X(v',0)$ $f$-value with $J'$ continues to the highest-excitation lines that we observe and is in perfect agreement with values predicted by the CSE model.
This decrease is effectively due to a decreasing Franck-Condon overlap of $b(v')$ and $X(0)$ vibrational wave functions with increasing centrifugal distortion.

\begin{figure}
  \centering
  \includegraphics{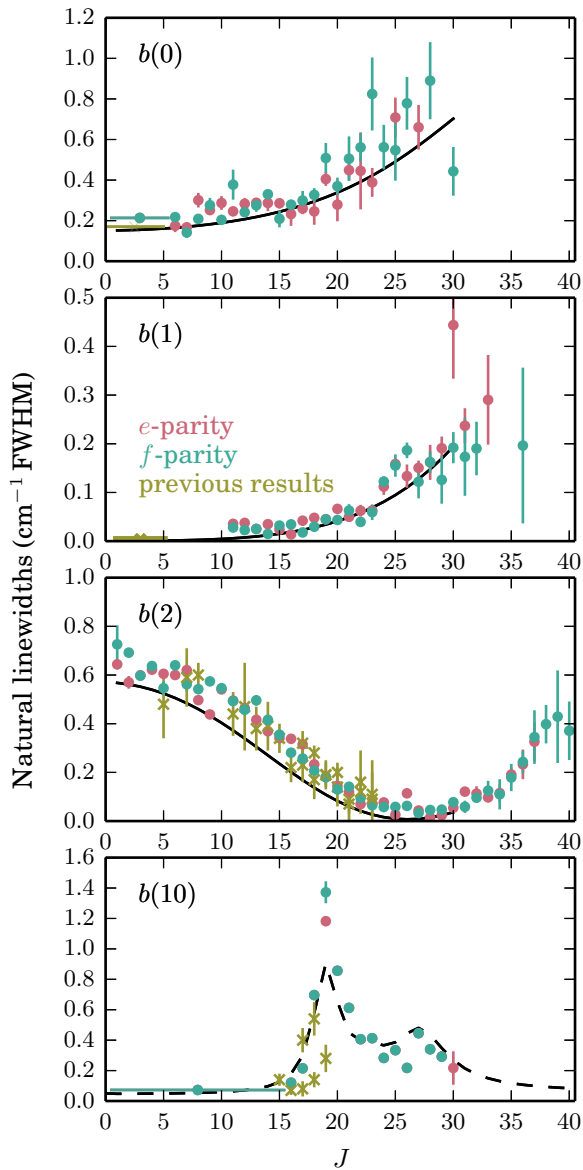}
  \caption{Natural linewidths of $e$- and $f$-parity excited-state levels accessed in our experiment as a function of their angular-momentum quantum number, $J'$, and with $1\sigma$ random fitting uncertainties (\emph{circles with error bars}). Some linewidths were analysed assuming $J'$-independent ranges (\emph{horizontal error bars}). Also shown are previously-measured linewidths~\cite{ubachs_etal1989,sprengers_etal2004b,sprengers_etal2005,stark_etal2005,stark_etal2008} (\emph{yellow lines and crosses}), and linewidths calculated by the CSE model (\emph{solid black curves}) and a two-level local interaction model (\emph{dashed black curve}).}
  \label{fig:compare widths}
\end{figure}Measured natural linewidths and comparable values from previous photoabsorption and resonantly-enhanced photoionisation experiments~\cite{ubachs_etal1989, sprengers_etal2005, stark_etal2005} are shown in Fig.~\ref{fig:compare widths}.
The widths of $b(0)$, $b(1)$, and $b(2)$ averaged over their $J\leq5$ levels have been previously deduced from laser-based lifetime or linewidth measurements~\cite{ubachs_etal1989,sprengers_etal2004b,sprengers_etal2005}.
The rotationally-resolved $J$-dependent broadening of $b(2)$ and $b(10)$ levels have been measured in synchrotron-based experiments~\cite{stark_etal2005, stark_etal2008}, and an increasing  $b(1)$ predissociation width with $J'$ has also been experimentally deduced~\cite{lewis_etal2005b,wu2012}.
Our newly-measured widths show good agreement with all reference data but with generally reduced scatter.
Two interesting new pieces of information are discussed below.

First, the decreasing $b(2)$ widths are now shown to pass through a minimum at $J\simeq 28$.
This complex behaviour is well-reproduced by the CSE model which includes a mechanism for predissociative line broadening by including unbound electronic states amongst its coupled channels~\cite{lewis_etal2005a,haverd2005}.
The critical interactions in this case are the spin-orbit coupling of $b(2)$ with vibrationally-bound levels of the $C\,{}^3\Pi_u$ state and their subsequent electronic interaction with the unbound continuum of the $C'\,{}^3\Pi_u$ state.
The dominant perturber of $b(2)$ is the $C(8)$ level which lies only 100\,cm$^{-1}$ lower in energy and has been previously identified in a photoabsorption spectrum~\cite{lewis_etal2008a} and found to have a linewidth of 18\,cm$^{-1}$ for $J'$ less than about 10,  despite the nominally spin-forbidden nature of this transition.
All other bound ${}^3\Pi_u$ states are too remote in energy to contribute significantly to the predissociation of $b(2)$~\cite{lewis_etal2008b} and the observed $J'$-dependence of its widths must then closely scale with the broader widths of $C(8)$.

\begin{figure}
  \centering
  \includegraphics{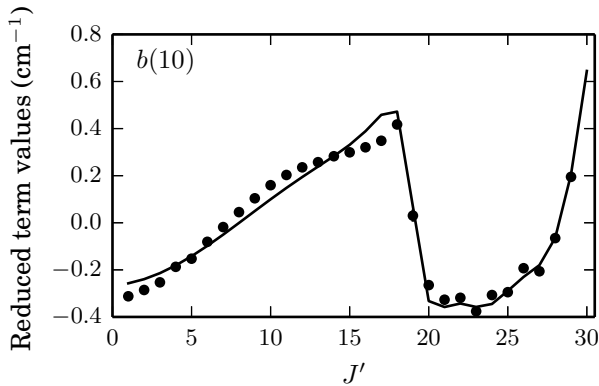}
  \caption{Experimental $f$-parity term values of $b(10)$ reduced by the subtraction of a cubic polynomial of best fit in terms of $J'(J'+1)$ (\emph{circles}). Also shown are reduced term values from the $b(10)$/${}^3\Pi_u$ interaction model (\emph{curve}).}
  \label{fig:compare b10 termres}
\end{figure}
Second, there is a sharp peak in the linewidths of $b(10)$ shown in Fig.~\ref{fig:compare widths}.
Increasing widths beginning around $J'=15$ were known from a poorer signal-to-noise-ratio room-temperature spectrum~\cite{stark_etal2008}, but are now better resolved and to higher-$J'$.
There is also a perturbation of $b(10)$ rotational energy levels near $J'=19$, as shown in Fig.~\ref{fig:compare b10 termres} as a 0.8\,cm$^{-1}$ deflection of its reduced term values.
The localised perturbation of $b(10)$ energies and widths indicates a level crossing with a predissociation-broadened level of ${}^3\Pi_u$ symmetry, as is
known to occur elsewhere in the N$_2$ spectrum~\cite{lewis_etal2008a}.

One candidate for the role of $b(10)$ perturber is the $v'=16$ level of the $C\,{}^3\Pi_u$ state, which has been observed for $J'\leq10$~\cite{lewis_etal2008a} and has a band origin only  80\,cm$^{-1}$ below that of $b(10)$.
However, a rotational constant calculated from the observed $C(16)$ levels, 1.153\,cm$^{-1}$~\cite{lewis_etal2008b}, is too low to cross the $b(10)$ term series where the observed perturbation peaks at $J\simeq 18$.
Alternatively, the $v=2$ level of the $G\,{}^3\Pi_u$ state has been observed~\cite{vanderkamp1994} to lie nearby, 340\,cm$^{-1}$ below $b(10)$, and undoubtedly has a larger rotational constant more characteristic of N$_2$ Rydberg levels, about $1.9$\,cm$^{-1}$.
A crossing between $G(2)$ and $b(10)$ is then conceivable.
The observed widths of $C(16)$ for $J'\leq10$ are less than $0.5$\,cm$^{-1}$\,FWHM, whereas $G(2)$ is predicted to be much broader, about 90\,cm$^{-1}$\,FWHM, by the CSE model of Lewis \emph{et al.}~\cite{lewis_etal2008b}.

To analyse the width and term value perturbation of $b(10)$ further we
defined a two-level model of $b(10)$ interacting by the spin-orbit
operator with a ${}^3\Pi_u$ level (including all triplet sublevels) and optimised its various parameters to match our experimental data.
This was done in an identical fashion to similar deperturbations of
N$_2$ ${}^3\Pi_u$/${}^1\Pi_u$ interactions by Lewis \emph{et al.}~\cite{lewis_etal2008a}.
Comparisons of experimental widths and reduced term values with this model are shown in Figs.~\ref{fig:compare widths} and \ref{fig:compare b10 termres} and find overall good agreement when adopting a  ${}^3\Pi_u$ state with a term origin of approximately 108\,150\,cm$^{-1}$, a rotational constant of 1.6\,cm$^{-1}$, and a spin-orbit splitting of 30\,cm$^{-1}$ (where the sign of the latter is unconstrained).
Two further model parameters are the strength of the $b(10)$ and ${}^3\Pi_u$ spin-orbit interaction, 7\,cm$^{-1}$, and the deperturbed predissociation widths of ${}^3\Pi_u$ levels.
Good agreement could only be found when assuming the latter increases linearly in term of $J'(J'+1)$ from 20\,cm$^{-1}$ at  $J'=18$, to  60\,cm$^{-1}$ at $J'=29$.
All of these deduced values are intermediate between those known or predicted for $C(16)$ and $G(2)$,~\cite{lewis_etal2008b,lewis_etal2008a,vanderkamp1994}, indicating that the perturber of $b(10)$ is an electronic admixture of the $C\,{}^3\Pi_u$ and $G\,{}^3\Pi_u$ states.
Indeed, the coupled-channels model of Lewis \emph{et al.}~\cite{lewis_etal2008b} predicts this, as well as a further significant admixture of the $F\,{}^3\Pi_u$ Rydberg state into the nominal $C(16)$ and $G(2)$ levels.

\section{Absorption spectra of CO A$-$X(0,0) and (1,0) bands}
\label{sec:COabsorption}

\begin{figure}
\resizebox{1\linewidth}{!}{\includegraphics{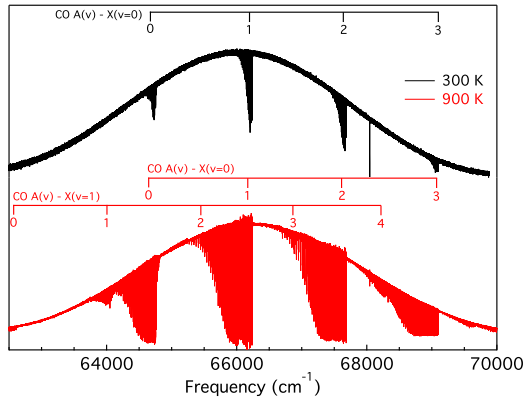}}
\caption{Overview spectrum of the CO \AX\ system including (0,0), (1,0), and (2,0) bands, and some hot bands. The top panel shows the spectrum which is measured at room temperature (300 K). In the bottom panel is the hot spectrum (900 K). The sharp absorption line in the upper spectrum at 68045.156 \wn\ is a xenon resonance line.}
\label{CO-overview}
\end{figure}

\begin{figure*}
\begin{center}
  \includegraphics[width=\textwidth]{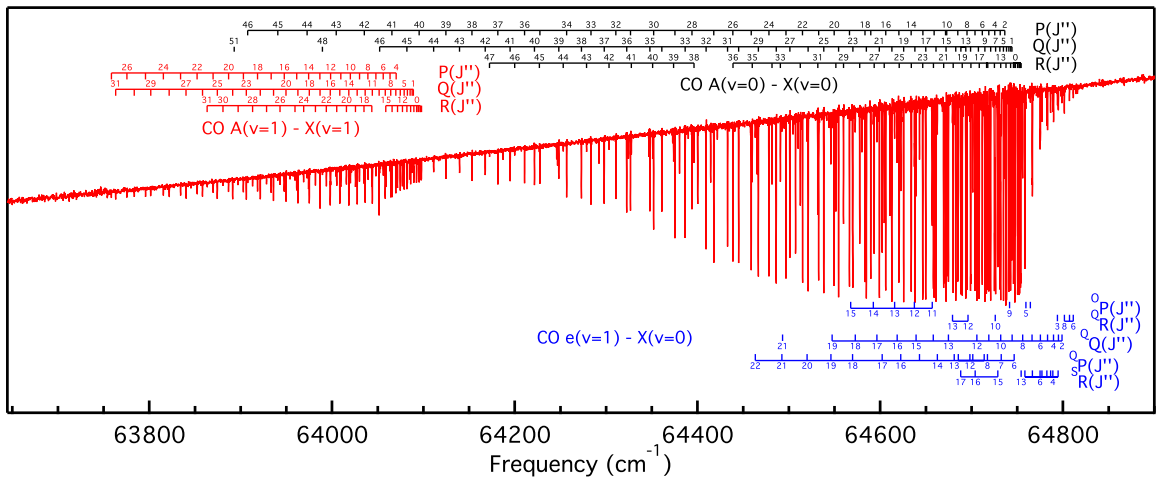}
\caption{(\emph{Preferably printed in two-column}) The spectrum of the CO \AX\ system in the range 63\,700 -- 64\,900 \wn\ measured using the hot cell in combination with the FTS-VUV. Rotational lines in the $A-X$ $(0,0)$ and $(1,1)$ bands are assigned by the sticks. Transitions to the \es\ perturber state are also assigned. The slope on the background continuum is due to the spectral profile of the undulator emission.}
  \label{Hotv0}
\end{center}
\end{figure*}

\begin{table*}
\begin{center}
\caption{Observed \change{high-$J$} transition frequencies (in vacuum \wn) of the CO \AX (0,0) and (1,0) bands obtained with the hot cell. Lower-$J$ transitions \change{are} listed in Ref.~\cite{niu2013}. The estimated uncertainty ($1\sigma$) is 0.02~\wn\  except for weak or blended lines.}
\label{AX00lines}
\scriptsize
\begin{tabular}
{cccccccc}
\hline
\hline
\multicolumn{1}{c}{} & \multicolumn{3}{c}{\AX(0,0)} & \multicolumn{1}{c}{} & \multicolumn{3}{c}{\AX(1,0)}	\\
\multicolumn{1}{c}{$J''$} & \multicolumn{1}{c}{R($J$)} & \multicolumn{1}{c}{Q($J$)} & \multicolumn{1}{c}{P($J$)} & \multicolumn{1}{c}{$J''$} & \multicolumn{1}{c}{R($J$)} & \multicolumn{1}{c}{Q($J$)} & \multicolumn{1}{c}{P($J$)}	\\
\hline
17	&		&		&	64589.64	&	21	&	66140.33	&		&	66005.05	\\
18	&		&		&	64583.37	&	22	&	66128.24	&		&	65986.78	\\
19	&	64689.77	&		&	64566.32	&	23	&	66115.44	&		&	65967.83	\\
20	&	64679.86	&		&	64549.58	&	24	&	66101.66	&	66023.95	&	65948.17	\\
21	&	64669.38	&		&	64532.53	&	25	&	66091.32	&	66006.31	&	65927.76	\\
22	&	64658.26	&	64584.85	&	64515.00	&	26	&	66074.03	&	65989.04	&	65906.37	\\
23	&	64646.44	&	64570.02	&	64496.90	&	27	&	66058.34	&	65970.28	&	65888.43	\\
24	&	64634.07	&	64554.44	&	64478.17	&	28	&	66042.09	&	65950.94	&	65863.57	\\
25	&	64620.76	&	64538.20	&	64458.80	&	29	&	66025.18	&	65929.54	&	65840.31	\\
26	&	64604.44	&	64521.15	&	64438.67	&	30	&	66007.60	&	65910.53	&	65816.49	\\
27	&	64593.32	&	64501.07	&	64417.88	&	31	&	65989.33	&	65889.04	&	65792.02	\\
28	&	64577.54	&	64486.13	&	64393.98	&	32	&	65970.28	&	65866.89	&	65766.89	\\
29	&	64559.78	&	64466.58	&	64375.28	&	33	&	65950.64	&	65843.98	&	65741.06	\\
30	&	64551.43	&	64445.05	&	64351.94	&	34	&	65930.24	&	65819.92	&	65714.55	\\
31	&	64531.64	&	64432.91	&	64326.63	&	35	&	65909.14	&	65798.34	&	65687.33	\\
32	&	64512.56	&	64409.35	&	64310.71	&	36	&	65887.33	&	65772.41	&	65659.43	\\
33	&	64490.76	&	64386.49	&	64283.40	&	37	&	65864.79	&	65746.65	&	65630.83	\\
34	&	64481.53	&	64360.95	&	64256.77	&	38	&	65841.53	&	65720.29	&	65601.52	\\
35	&	64459.85	&	64347.94	&	64227.48	&	39	&	65817.50	&	65693.27	&	65571.51	\\
36	&	64438.85	&	64322.50	&	64210.71	&	40	&	65793.18	&	65665.55	&	65540.79	\\
37	&		&	64297.87	&	64181.54	&	41	&	65767.61	&	65637.11	&	65509.30	\\
38	&	64396.16	&	64272.94	&	64153.19	&	42	&	65741.50	&	65607.95	&	65477.54	\\
39	&	64373.87	&	64248.16	&	64124.55	&	43	&	65714.55	&	65577.99	&	65444.54	\\
40	&	64350.93	&	64221.59	&	64095.36	&	44	&	65686.94	&	65548.84	&	65410.97	\\
41	&	64327.65	&	64194.95	&	64065.55	&	45	&	65658.53	&	65516.48	&	65376.69	\\
42	&	64303.14	&	64167.66	&	64035.39	&	46	&	65628.96	&	65484.32	&	65341.67	\\
43	&	64278.62	&	64139.56	&	64004.55	&	47	&	65600.90	&	65451.21	&	65305.83	\\
44	&	64253.03	&	64111.24	&	63972.67	&	48	&	65569.47	&	65418.96	&	65268.92	\\
45	&	64226.85	&	64081.88	&	63940.64	&	49	&	65537.24	&	65383.97	&		\\
46	&	64199.92	&	64052.09	&	63907.71	&	50	&		&	65348.14	&		\\
47	&	64172.37	&		&		&	51	&		&	65318.08	&		\\
48	&		&	63989.48	&		&	52	&		&	65277.96	&		\\
51	&		&	63892.98	&		&	53	&	65402.39	&		&		\\
\hline
\hline
\end{tabular}
\end{center}
\end{table*} 
The novel hot cell configuration was employed for the further investigation of the \AX\ system of CO for the lowest vibrational bands in the excited state. Figure~\ref{CO-overview} shows an overview spectrum of some bands recorded at 300 K and 900 K. With the higher gas temperature, \change{the} rotational envelope of each band includes higher $J$-quantum numbers and hot bands appear \change{that originate} from $X^1\Sigma^+(v''=1)$. Figure~\ref{Hotv0} displays a more detailed spectrum of the $(0,0)$ and $(1,1)$ bands of the \AX\ system of CO. Note that the strongest transitions in Figs.~\ref{CO-overview} and \ref{Hotv0} are saturated. A number of spectra were recorded at various gas densities so that transition frequencies for all lines could be analysed in unsaturated recordings. The observed \change{high-$J$} transition frequencies in the \AX $(0,0)$ and $(1,0)$ bands are collected in Table~\ref{AX00lines}. While in a previous room temperature study of the same bands their rotational progression could be followed \change{up} to $J=21$ and $J=23$, respectively~\cite{niu2013}, the present spectrum reveals lines up to $J=51$ and $J=53$ for the two bands. \change{Accurate transition frequencies for low-$J$ transitions were already given in Ref.\cite{niu2013}.}

\begin{table}
\begin{center}
\begin{small}
\caption{Observed transition frequencies (in vacuum \wn) of the CO \eX (1,0) band obtained with the hot cell. The estimated uncertainty ($1\sigma$) is 0.02~\wn\  except for weak or blended lines.}
\label{eX10lines}
\scriptsize
\begin{tabular}
{cccccc}
\hline
\hline
\\[-1ex]
\multicolumn{1}{c}{} & \multicolumn{2}{c}{\change{$F_1$}} & \multicolumn{1}{c}{\change{$F_2$}} & \multicolumn{2}{c}{\change{$F_3$}}	\\
\hline
\\[-1ex]
\multicolumn{1}{c}{$J''$} & \multicolumn{1}{c}{\change{$^Q$R($J$)}} & \multicolumn{1}{c}{\change{$^O$P($J$)}} & \multicolumn{1}{c}{\change{$^Q$Q($J$)}} & \multicolumn{1}{c}{\change{$^S$R($J$)}} & \multicolumn{1}{c}{\change{$^Q$P($J$)}}	\\
\hline
\\[-1ex]
2	&		&		&	64798.89	&		&		\\
3	&	64794.01	&		&	64794.98	&		&		\\
4	&		&		&	64789.71	&	64788.84	&		\\
5	&		&	64759.37	&	64783.15	&	64782.45	&		\\
6	&	64811.33	&		&	64775.34	&	64774.92	&	64746.55	\\
7	&	64807.10	&	64764.30	&	64766.22	&	64766.60	&	64732.49	\\
8	&	64801.53	&		&	64755.89	&	64758.41	&	64717.25	\\
9	&		&	64741.55	&	64744.31	&	64794.65	&	64701.25	\\
10	&	64726.00	&		&	64731.92	&	64786.42	&	64685.41	\\
11	&		&	64656.88	&	64718.79	&	64776.94	&	64713.88	\\
12	&	64696.16	&	64637.39	&	64705.72	&		&	64698.09	\\
13	&	64679.05	&	64615.71	&	64674.65	&	64754.31	&	64680.93	\\
14	&		&	64592.43	&	64658.00	&		&	64662.50	\\
15	&		&	64567.68	&	64639.20	&	64728.76	&	64642.97	\\
16	&		&		&	64618.52	&	64703.94	&	64622.54	\\
17	&		&		&	64596.30	&	64688.01	&	64602.10	\\
18	&		&		&	64572.61	&		&	64569.64	\\
19	&		&		&	64547.41	&		&	64545.95	\\
20	&		&		&		&		&	64520.00	\\
21	&		&		&	64493.11	&		&	64492.39	\\
22	&		&		&		&		&	64463.22	\\
\hline
\hline
\end{tabular}
\end{small}
\end{center}
\end{table}

\begin{table*}
\begin{center}
\caption{Observed transition frequencies (in vacuum \wn) of the CO \dX (4,0) band obtained with the hot cell. The estimated uncertainty ($1\sigma$) is 0.02~\wn\  except for weak or blended lines.}
\label{dX40lines}
\scriptsize
\begin{tabular}
{cccccccc}
\hline
\hline
\\[-1ex]
\multicolumn{1}{c}{} & \multicolumn{3}{c}{\change{$F_1$}} & \multicolumn{3}{c}{\change{$F_2$}} & \multicolumn{1}{c}{\change{$F_3$}}	\\
\hline
\\[-1ex]
\multicolumn{1}{c}{$J''$} & \multicolumn{1}{c}{R($J$)} & \multicolumn{1}{c}{Q($J$)} & \multicolumn{1}{c}{P($J$)} & \multicolumn{1}{c}{R($J$)} & \multicolumn{1}{c}{Q($J$)} & \multicolumn{1}{c}{P($J$)} & \multicolumn{1}{c}{Q($J$)}	\\
\hline
\\[-1ex]
26	&	64610.70	&	64544.08	&		&		&		&		&		\\
27	&		&	64507.33	&		&		&		&		&		\\
28	&		&		&	64400.24	&		&		&		&		\\
29	&		&		&		&	64579.94	&	64504.55	&		&		\\
30	&		&		&		&	64532.81	&	64465.22	&	64389.87	&		\\
31	&		&		&		&	64495.05	&	64414.40	&	64346.77	&		\\
32	&		&		&		&		&	64372.81	&	64292.14	&	64464.02	\\
33	&		&		&		&	64510.13	&	64421.74	&		&		\\
34	&		&		&		&	64458.06	&	64380.33	&		&		\\
35	&		&		&		&		&	64324.46	&	64246.83	&		\\
36	&		&		&		&		&		&	64187.21	&		\\
\hline
\hline
\end{tabular}
\end{center}
\end{table*}
\begin{table*}
\begin{center}
\begin{small}
\caption{Observed transition frequencies (in vacuum \wn) of the CO \dX (5,0) band obtained with the hot cell. The estimated uncertainty ($1\sigma$) is 0.02~\wn\  except for weak or blended lines.}
\label{dX50lines}
\scriptsize
\begin{tabular}
{ccccccc}
\hline
\hline
\\[-1ex]
\multicolumn{1}{c}{} & \multicolumn{3}{c}{\change{$F_1$}} & \multicolumn{3}{c}{\change{$F_2$}}\\
\hline
\\[-1ex]
\multicolumn{1}{c}{$J''$} & \multicolumn{1}{c}{R($J$)} & \multicolumn{1}{c}{Q($J$)} & \multicolumn{1}{c}{P($J$)} & \multicolumn{1}{c}{R($J$)} & \multicolumn{1}{c}{Q($J$)} & \multicolumn{1}{c}{P($J$)}\\
\hline
\\[-1ex]
0	&		&		&		&	66210.18	&		&		\\
1	&		&		&		&	66211.77	&		&		\\
2	&	66176.59	&		&		&	66212.17	&	66204.04	&	66198.61	\\
3	&	66174.74	&		&		&	66211.19	&	66200.58	&		\\
4	&	66171.62	&		&	66149.75	&	66209.16	&	66195.87	&	66185.22	\\
5	&		&		&	66140.16	&		&	66189.94	&		\\
6	&	66160.90	&	66143.98	&		&	66201.11	&	66182.70	&	66166.87	\\
7	&		&		&	66116.88	&	66195.03	&	66174.17	&	66155.80	\\
8	&		&		&		&	66187.68	&	66164.30	&	66143.38	\\
9	&		&		&		&	66178.92	&	66153.03	&	66129.72	\\
10	&		&	66095.82	&		&	66168.86	&	66140.42	&	66114.67	\\
11	&		&		&	66053.23	&	66157.25	&	66126.61	&	66098.22	\\
12	&		&		&		&	66144.58	&	66111.32	&	66080.43	\\
13	&		&		&		&	66130.29	&	66094.65	&	66061.40	\\
14	&		&		&		&	66114.67	&	66076.58	&	66040.93	\\
15	&		&		&		&	66097.74	&	66057.17	&	66019.02	\\
16	&		&		&		&		&	66036.44	&	65995.79	\\
17	&		&		&		&		&	66014.27	&	65971.17	\\
\hline
\hline
\end{tabular}
\end{small}
\end{center}
\end{table*}
\begin{table}
\begin{center}
\begin{small}
\caption{Observed transition frequencies (in vacuum \wn) of other CO electronic-vibrational bands obtained with the hot cell. The estimated uncertainty ($1\sigma$) is 0.02~\wn\  except for weak or blended lines.}
\label{otherslines}

\begin{tabular}
  {cc}
  \hline
  \hline
  \\[-1ex]
  \begin{tabular}{cc}
    \multicolumn{2}{c}{\DX(1,0)} \\[1ex]
    $^P$P(26) &	65916.49\\
    $^P$P(27) &	65875.99\\
    $^Q$Q(24) &	66050.52\\
    $^Q$Q(25) &	66016.10\\
    $^Q$Q(26) &	65979.36\\
    $^R$R(25) &	66078.85\\
    $^R$R(27) &	66011.16\\
  \end{tabular}&
  \begin{tabular}{cc}
    \multicolumn{2}{c}{\aX(9,0)}\\[1ex]
    $^P$Q1(39)                  &	64245.88\\[2ex]
    \hline\\[-1ex]
    \multicolumn{2}{c}{\dX(5,1)}\\[1ex]
    $^R$Q31(3)                  &	64057.44\\
    $^R$Q31(4)                  &	64052.92\\[2ex]
    \hline\\[-1ex]
    \multicolumn{2}{c}{\IX(2,0)}\\[1ex]
    $^Q$Q(34)                    &	65837.46\\
    $^Q$Q(35)                    &	65785.95\\
  \end{tabular}\\
  \noalign{\medskip}
  \hline\hline\\
\end{tabular}

\end{small}
\end{center}
\end{table}

\begin{table}
\begin{center}
\begin{small}
\setlength{\tabcolsep}{4pt}
\caption{\change{The updated molecular constants for the \As\ (\0) and (\1) states of \CO\ and the perturber states \ds\ (\4) and \Ds\ (\1) following from the perturbation analysis. In cases where an uncertainty is specified in () parentheses the value was determined from the fit; in cases where this is not specified a value was taken from literature. For the \As\ states, the $T_v$ and $B$ parameters are fixed to the previous work\cite{niu2013}, because these parameters are predominantly determined by the low-$J$ transition frequencies. All values in vacuum \wn.}}
\label{COconstants}
\begin{tabular}
{lrlr}
\hline
\hline
\\[-1ex]
\multicolumn{2}{c}{\As(\0)} &\multicolumn{2}{c}{\As(\1)}  \\
\hline
\\[-1ex]
$T_v$	&	64746.762	&$T_v$	&	66228.801	\\
$B$	&	1.604069	&$B$	&	1.58126	\\
$q$ ($\times 10^5$)	&	1.4	(4)	&$q$ ($\times 10^5$)	&	-2.5	(4)	\\
$D$ ($\times 10^6$)	&	7.352	(5)	&$D$ ($\times 10^6$)	&	7.438	(4)	\\
$H$ ($\times 10^{12}$)	&	-8	(2)	&$H$ ($\times 10^{12}$)	&	-15	(1)	\\
\hline
\\[-1ex]
\multicolumn{2}{c}{\ds(\4)}	&\multicolumn{2}{c}{\Ds(\1)}			\\
\hline
\\[-1ex]
$T_v$	&	65101.90	(3)	&	$T_v$	&	66462.11	(9)	\\
$B$	&	1.23381	(8)	&	$B$	&	1.2373	(3)	\\
$A$	&	-16.52	(1)	&		&			\\
$\lambda$&	1.15	(3)	&		&			\\
$\gamma$ ($\times 10^3$)	&	-8.54		&		&			\\
$D$ ($\times 10^6$)	&	6.80	(6)	&	$D$ ($\times 10^6$)	&	8.8	(4)	\\
$H$ ($\times 10^{12}$)	&	-0.8		&	$H$ ($\times 10^{12}$)	&	-0.3		\\
$A_D$ ($\times 10^4$)	&	-1		&		&			\\
$\eta_0$	&	-21.72	(1)	&	$\xi_0$	&	0.040		\\
$\eta_1$	&			&	$\xi_1$	&	0.077	(1)	\\
\hline
\hline
\end{tabular}
\end{small}
\end{center}
\end{table}
In the observed region, \change{between} $63\,500 - 67\,500$ \wn, many lines were observed that excite perturber states of the \As\ ($v=0$) and ($v=1$) levels, and are examined in the present study.
Lines pertaining to the \eX\ (1,0) band, clearly visible in Fig.~\ref{Hotv0}, are listed in Table~\ref{eX10lines}. Data for the \eX\ (1,0) band had previously been reported by Simmons and Tilford~\cite{Simmons1971} and, at higher accuracy and also up to $J=22$, by Lefloch~\emph{et al.}~\cite{Lefloch1987}. On average the data are offset by $0.04$ \wn \change{with respect to present values}, which is within the error margins claimed in Ref.~\cite{Lefloch1987}. Term values of the \es\ ($v=1$) level can also be obtained via measurement of lines in the \Be\ system, with observations of the (0,1) band~\cite{choe1989} at accuracies in the range $0.001 - 0.02$ \wn\ combined with the measurements of the \BX(0,0) band, accurate at $0.003$ \wn~\cite{drabbels1993}. In comparison with the present data set the overall offset on the term values is within 0.015 \wn, which is well within the quoted uncertainties.

Observed lines associated with the \dX\ system are listed in Table~\ref{dX40lines} for the (4,0) band, observed for rotational angular momenta of $J=26-36$, and in Table~\ref{dX50lines} for the (5,0) band, with observation of $J=0-17$. In an investigation by Herzberg~\emph{et al.}~\cite{herzberg1970} rotational levels up to $J=22$ were observed in both bands at low accuracy. The observations in the \dX (4,0) band were superseded by those of Lefloch~\emph{et al.}~\cite{Lefloch1987} at higher accuracy. Comparison with the latter and the present data set yields agreement within $0.04$ \wn, hence within the quoted error margin of $0.06$ \wn\ in Ref.~\cite{Lefloch1987}.

\change{The \dX (5,0) band had been investigated by VUV laser-induced fluorescence measurements~\cite{duplessis2007}.
Accurate data on this band were also reported from classical spectroscopic studies by Lefloch~\cite{Lefloch-thesis}; for this set the agreement with the present data is within $0.02$ \wn. The \ds\ ($v=5$) levels were also observed in emission in the \Bd(0,5) band~\cite{choe1989}.
In the present study the $F_1$ and $F_2$ fine structure components in the \ds\ state were observed,  while in the study of Choe~\emph{et al.}~\cite{choe1989} the $F_3$ components were seen as a result of different intensity borrowing,
In addition two lines in the \dX (5,1) band were observed and listed in Table~\ref{otherslines}.}


Additional lines observed were assigned to the \DX(1,0) band, the \IX(2,0) band, and the \aX(9,0) band, and listed in Table~\ref{otherslines}.
Some of these transitions probing perturber states were observed previously by Lefloch~\emph{et al.}~\cite{Lefloch1987, Lefloch-thesis}, although not all, and at a lower accuracy of 0.06 \wn.
Herzberg~\emph{et al.} observed states up to $J=22$ in the \IX(2,0) band~\cite{herzberg1966}. Despite the fact that information on the intermediate $J$-levels is missing, an unambiguous assignment of the transitions originating in $J=34-35$ could nevertheless be made based on the perturbation patterns.
The same holds for the newly observed lines in the \DX(1,0) band, for which rotational lines up to $J=17$ were observed in the past~\cite{simmons1966}, and for which new lines originating from $J=24-26$ are found.

A reiteration of a previous deperturbation analysis for the \As~($v=0$) and ($v=1$) states~\cite{niu2013} is performed including the additional data points for high-$J$ levels.
A comprehensive fit was performed based on the diagonalisation of a series of matrices containing $J$-dependent deperturbed energy levels and interaction energies of multiple states. The entire set of experimental \AX (0,0) and (1,0) lines were reproduced and all lines exciting perturber states.
The form of these perturbation matrices is kept the same as that defined in Table 6 of Ref.~\cite{niu2013}\change{, keeping the same labels for the parameters}.
For the \As\ states, the $T_v$ and $B$ parameters are fixed to the previous work\cite{niu2013}, because these parameters are predominantly determined by the low-$J$ transition frequencies.
Most of the molecular constants resulting from this procedure did not undergo
a significant change except for the values pertaining to the states \Ds($v=1$) and \ds($v=4$). These values are listed in Table~\ref{COconstants}. The main difference for the \ds\ state entails the inclusion of a quartic centrifugal distortion D and a spin-spin coupling constant, $\lambda$. Values for the \Ds\ state were previously kept constant but are now optimized in the present fit.

The rotational temperature $T_{rot} =  927 \pm 20$ K is determined by fitting the transition intensities of transitions with different $J$--quantum numbers, assuming a Boltzmann distribution of ground state populations. This fitting considers perturber states borrowing intensity from the \As-\Xs transitions. The vibrational temperature, $T_{vib} \sim 845$ K, is calculated by comparing the intensities of the strong \AX$(1,0)$ band and the weak $(1,1)$ hot band. For this analysis pressure saturation effects must be considered that prevent a direct comparison of intensities of these two vibrational bands with very different cross sections. Instead, spectra recorded at different column densities are used, also involving a comparison with the \AX(0,0) band of intermediate strength.  Further, Franck-Condon factors of the $(1,0)$ and $(1,1)$ vibrational transitions must be considered, which are taken from Ref.~\cite{beegle1999}.
The kinetic temperature, associated with Doppler broadening, is determined at $T_{kin} \sim 900$ K.

\section{Conclusion}

Vacuum-ultraviolet photoabsorption spectra of N$_2$ and CO were recorded at 930\,K using a heated free-flowing gas cell and the Fourier-transform spectrometer end station of the DESIRS beamline at the SOLEIL synchrotron.
This novel setup allowed  for the measurement of rotational transitions with angular-momentum quantum numbers, $J'$, as high as 51 and also from the first excited ground state vibrational level, which is well beyond the limit of room-temperature experiments.
The high-resolution spectrometer permitted quantification of rotationally-resolved transition energies, $f$-values, and predissociation broadening for many vibrational bands.

In CO, we deduce new high-$J'$ level energies for the upper states of the $A\,{}^1\Pi - X\,{}^1\Sigma^+(v',v''=0)$ bands with $v'=0$ and 1, as well as observe new forbidden transitions to levels of the $e\,{}^3\Sigma^-$, $d\,{}^3\Delta$, $D\,{}^1\Delta$ and $a'\,{}^3\Sigma^+$ states.
The forbidden transitions appear due to intensity-borrowing from the $A-X$ bands and the new data permitted an improved estimate of molecular parameters describing the forbidden levels and their perturbing interactions.

We measure new level energies, $f$-values, and predissociation linewidths of the N$_2$ bands $b\,{}^1\Pi_u- X\,{}^1\Sigma^+_u(v',v''=0)$  for $v'=0,1,2$, and 10.
These verify the high-$J'$ predictions of a CSE model which was constructed with respect to room-temperature experimental data.
This validates the use of photodissociation cross sections calculated from this model in atmospheric or astrophysical applications at high temperatures.
No forbidden levels are observed for the case of N$_2$.
Instead, the $J'$-dependent widths of $b(v')$ provide new indirect information on an interacting dissociative manifold of ${}^3\Pi$ levels.
The analysis of perturbed $b(10)$ level energies and linewidths permitted the characterisation of its level crossing with a ${}^3\Pi_u$ level of mixed electronic character.

There is some indication that for both target molecules the rotational and vibrational temperatures are not identical, 930 and 800--830\,K, respectively.
The incomplete equilibration of vibrational and rotational excitation will not affect our conclusions regarding the perturbation of high $J'$ levels but introduces uncertainty into any determination of hot-band absolute $f$-values.

The present measurements of high-temperature cross sections and line lists have a direct application to the study of astrophysical environments and planetary atmospheres.
It also provides detailed extra information on the underlying electronic states of the molecules and their non-Born-Oppenheimer interactions.
This information, when incorporated with the larger experimental record, is necessary for constraining predictive models of the photoabsorbing and dissociating excited states; and will allow for improvements to the N$_2$ CSE model, and similar theoretical developments for the case of CO.

\section*{Acknowledgements}
This work is financially supported by the Dutch Astrochemistry Network of the Netherlands Foundation for Scientific Research (NWO).
We wish to thank JF Gil, technician on the DESIRS beam line for the mechanical conception of the hot cell. We are grateful to the general and technical staff of SOLEIL for providing beam time under project n$^o20120653$.



\end{document}